\documentclass{emulateapj}
\newcommand{\etal}{et al.\ }
\newcommand{\etalb}{et al.}
\newcommand{\beq}{\begin{equation}}
\newcommand{\beqa}{\begin{eqnarray}}
\newcommand{\eeq}{\end{equation}}
\newcommand{\eeqa}{\end{eqnarray}}

\newcommand{\br}{{\bf r}}
\newcommand{\bk}{{\bf k}}
\newcommand{\Lya}{Ly$\alpha$~}
\newcommand{\Lyb}{Ly$\beta$~}
\newcommand{\td}{{\tilde{\delta}}}

\usepackage{apjfonts}
 
\begin{document}
\title{Detecting the Earliest Galaxies Through Two New Sources of 21cm 
Fluctuations}

\author{Rennan Barkana\altaffilmark{1} \& Abraham
  Loeb\altaffilmark{2}}

\altaffiltext{1} {School of Physics and Astronomy, The Raymond and 
Beverly Sackler Faculty of Exact Sciences, Tel Aviv University, Tel
Aviv 69978, ISRAEL; barkana@wise.tau.ac.il}

\altaffiltext{2} {Astronomy Department, Harvard University, 60 
Garden Street, Cambridge, MA 02138; aloeb@cfa.harvard.edu}

\begin{abstract}

The first galaxies that formed at a redshift $z\sim 20$--30 emitted
continuum photons with energies between the Ly$\alpha$ and Lyman limit
wavelengths of hydrogen, to which the neutral universe was transparent
except at the Lyman-series resonances. As these photons redshifted or
scattered into the Ly$\alpha$ resonance they coupled the spin
temperature of the 21cm transition of hydrogen to the gas temperature,
allowing it to deviate from the microwave background temperature. We
show that the fluctuations in the radiation emitted by the first
galaxies produced strong fluctuations in the 21cm flux before the
Ly$\alpha$ coupling became saturated. The fluctuations were caused by
biased inhomogeneities in the density of galaxies, along with Poisson
fluctuations in the number of galaxies. Observing the power-spectra of
these two sources would probe the number density of the earliest
galaxies and the typical mass of their host dark matter halos. The
enhanced amplitude of the 21cm fluctuations from the era of \Lya
coupling improves considerably the practical prospects for their
detection.

\end{abstract}

\keywords{galaxies: high-redshift, cosmology: theory, galaxies:
formation}

\section{Introduction}

The reionization history of cosmic hydrogen, left over from the big
bang, provides crucial fossil evidence for when the first stars and
black holes formed in the infant universe \citep{BL01}. The hyperfine
spin-flip transition of neutral hydrogen (\ion{H}{1}) at a wavelength
of 21 cm is potentially the most promising probe of the cosmic gas
before reionization ended. Observations of this line at a wavelength
of $21\times (1+z)$ cm can be used to slice the universe as a function
of redshift $z$ and obtain a three-dimensional map of the diffuse
\ion{H}{1} distribution in it \citep{Hogan}. The 21cm signal vanished 
at redshifts $z\ga 200$, when the small residual fraction of free
electrons after cosmological recombination kept the gas kinetic
temperature, $T_k$, close to the temperature of the cosmic microwave
background (CMB), $T_\gamma$.  Subsequently, between $200\ga z\ga 30$
the gas cooled adiabatically, faster than the CMB, and atomic
collisions kept the spin temperature $T_s$ of the hyperfine level
population below $T_\gamma$, so that the gas appeared in absorption
\citep{Scott}. Primordial density inhomogeneities imprinted a
three-dimensional power-spectrum of 21cm flux fluctuations on scales
down to $\la 10$ comoving kpc, making it the richest data set on the
sky \citep{Loeb04}. As the Hubble expansion continued to rarefy the
gas, radiative coupling of $T_s$ to $T_\gamma$ dominated over
collisional coupling of $T_s$ to $T_k$ and the 21cm signal began to
diminish.

However, as soon as the first galaxies appeared, the UV photons they
produced between the Ly$\alpha$ and Lyman limit wavelengths propagated
freely through the universe, redshifted or scattered into the
Ly$\alpha$ resonance, and coupled $T_s$ and $T_k$ once again through
the Wouthuysen-Field \citep{Wout, Field} effect by which the two
hyperfine states are mixed through the absorption and re-emission of a
Ly$\alpha$ photon \citep{Madau, Ciardi}. Emission of UV photons above
the Lyman limit by the same galaxies initiated the process of
reionization by creating ionized bubbles in the neutral gas around
these galaxies. While UV photons were blocked at the neutral boundary
of the bubbles, X-ray photons propagated farther into the bulk of the
intergalactic gas and heated $T_k$ above $T_\gamma$ throughout the
universe. Once $T_s$ grew larger than $T_\gamma$, the gas appeared in
21cm emission with a brightness temperature that was asymptotically
independent of $T_s$ and of $T_k$.  The ionized bubbles imprinted a
knee in the power-spectrum of 21cm fluctuations \citep{Zalda04}, which
traced the topology of \ion{H}{1} as the ionized bubbles percolated
and completed the process of reionization \citep{Fur04}.

Since the first galaxies represented rare peaks in the cosmic density
field, their spatial distribution possessed unusually large
fluctuations on large scales \citep{Bar04a}. In this paper we show
that the resulting large-scale bias of the first galaxies produced
substantial 21cm fluctuations during the intermediate epoch when the
first Ly$\alpha$ photons coupled $T_s$ to $T_k$. Since a relatively
small number of galaxies contributed to the flux seen at any given
point, Poisson fluctuations were significant as well, and we show that
they produced correlated 21cm fluctuations. Owing to the rapid
evolution in the density of galaxies with cosmic time, there is a
sharp decline in the abundance of sources with distance (corresponding
to retarded cosmic time for the source emission due to the light
travel time) around each point in the intergalactic medium.
Ultimately, there is an effective horizon around each point out to
which sources may contribute to 21cm fluctuations. In addition, the
atomic physics of resonant scattering by hydrogen fixes different
horizon distances for photons emitted in various wavelength intervals
between \Lya and the Lyman limit; these horizons are set by the
requirement that a given photon must redshift or scatter into the \Lya
resonance at the point of interest. We incorporate all of these
ingredients into our calculation.

The outline of the paper is as follows. In \S~2 we summarize the basic
equations that describe the evolution of the spin temperature. In \S~3
we review the angular anisotropy of the 21cm power spectrum due to
peculiar velocities, as derived in our previous paper \citep{BL04}. We
then calculate the flux of \Lya photons produced by the earliest
galaxies (\S~4) and its fluctuations (\S~5). Finally, we demonstrate
how the resulting power spectrum of fluctuations in the 21cm flux
(\S~6) can be used to study the properties of the earliest galaxies
(\S~7), and briefly summarize our main results (\S~8). Throughout the
paper we assume the concordance set of cosmological parameters
\citep{WMAP}.

\section{Spin Temperature History}

The \ion{H}{1} spin temperature, $T_s$, is defined through the ratio
between the number densities of hydrogen atoms in the excited and
ground state levels, ${n_1/ n_0}=(g_1/ g_0)\exp\left\{-{T_\star/
T_s}\right\}$, where $(g_1/g_0)=3$ is the ratio of the spin degeneracy
factors of the levels, and $T_\star=0.0682$K. When the spin
temperature is smaller than the CMB temperature, neutral hydrogen
atoms absorb CMB photons. The resonant 21cm absorption reduces the
brightness temperature of the CMB by \citep{Scott, Madau}
\begin{equation}
T_b =\tau \left({T_s-T_{\gamma}\over 1+z}\right)\ ,
\end{equation}
where the optical depth for resonant $\lambda=21$cm absorption is,
\begin{equation}
\tau= \frac {3c\lambda^2hA_{10}n_{\rm H}} {32 \pi k_B T_s\, (1+z)\, 
(dv_r/dr)}\ , \label{eq:tau}
\end{equation}
where $n_H$ is the number density of hydrogen, $A_{10}=2.85\times
10^{-15}~{\rm s^{-1}}$ is the spontaneous emission coefficient, and
$dv_r/dr$ is the gradient of the radial velocity along the line of
sight with $v_r$ being the physical radial velocity and $r$ the
comoving distance; on average $dv_r/dr = H(z)/ (1+z)$ where $H$ is the
Hubble parameter. The velocity gradient term arises since the 21cm
scattering cross-section has a fixed thermal width which translates
through the redshift factor $(1+v_r/c)$ to a fixed interval in
velocity \citep{Sobolev60}. In this paper we consider the epoch long
before complete reionization, and we assume that the hydrogen gas is
almost entirely neutral.

The mean brightness temperature offset on the sky at redshift $z$ is 
\begin{equation}
T_b = 28\, {\rm mK}\, \left( \frac{\Omega_b h} {.033} \right) \left(
\frac{\Omega_m} {.27} \right)^{-1 /2} \left({{1+z}\over{10}}\right)^{1/2} 
\left({{T_s - T_{\gamma}}\over {T_s}}\right)\ ,
\end{equation}
where we have substituted the concordance values for the cosmological
parameters $\Omega_b$, $h$, and $\Omega_m$.  The spin temperature
itself is coupled to $T_k$ through the spin-flip transition, which can
be excited by atomic collisions or by the absorption of \Lya
photons. Assuming that the photons near \Lya are redistributed in
frequency according to the thermal velocities of the atoms (i.e., the
spectrum is thermalized according to $T_k$), the combination that
appears in $T_b$ becomes \citep{Field}
\begin{equation}
{T_s - T_{\gamma} \over T_s} = {x_{\rm tot}\over 1+ x_{\rm tot}}
\left(1 - {T_{\gamma}\over T_k} \right)\ , \label{eq:combo}
\end{equation}
where $x_{\rm tot} = x_{\alpha} + x_c$ is the sum of the radiative and
collisional threshold parameters. These parameters are
\begin{equation}
x_{\alpha} = {{4 P_{\alpha} T_\star}\over {27 A_{10} T_{\gamma}}}\ ,
\end{equation}
and 
\begin{equation}
x_c = {{4 \kappa_{1-0}(T_k)\, n_H T_\star}\over {3 A_{10}
T_{\gamma}}}\ ,\ \end{equation} where $P_{\alpha}$ is the \Lya
scattering rate which is proportional to the \Lya intensity, and
$\kappa_{1-0}(T_k)$ is tabulated as a function of $T_k$
\citep{AD, Zyg}. Note that we have adopted a modified notation compared
to the $y_\alpha$ and $y_c$ previously used \citep{Field, Madau},
defining $x_\alpha \equiv y_\alpha(T_\alpha/T_\gamma)$ and $x_c
\equiv y_c(T_k/T_\gamma)$. The coupling of the spin temperature
to the gas temperature becomes substantial when $x_{\rm tot} \ga 1$;
in particular, $x_{\alpha} = 1$ defines the thermalization rate
\citep{Madau} of $P_{\alpha}$, which is $7.69 \times 10^{-12} (1+z)/10$ 
photons per second.

\section{The Separation of Powers}

\label{sec:sep}

Although the mean 21cm emission or absorption is difficult to measure
due to bright foregrounds, the unique character of the fluctuations in
$T_b$ allows for a much easier extraction of the signal
\citep{Shaver, Zalda04, Miguel1, Miguel2, Santos}. During the era of 
initial \Lya coupling, fluctuations in $T_b$ are sourced by
fluctuations in the density ($\delta$), temperature ($\delta_{T_k}$)
and radial velocity gradient of the gas as well as the \Lya flux
through $x_{\alpha}$ ($\delta_{x_\alpha}$). The fluctuations in the
neutral fraction of the gas become important only later, when the
volume filling fraction of ionized bubbles is substantial.

The above sources of fluctuations are all isotropic except for the
velocity gradient term. Since this term involves radial projections,
it inserts an anisotropy into the power spectrum \citep{kaiser, tozzi,
Indian}. The total power spectrum can be written as \citep{BL04} \beqa
P_{T_b}(\bk) & = & \mu^4 P_{\delta}(k) + 2 \mu^2 \left[ \beta
P_{\delta}(k) + \frac{x_{\alpha}} {\tilde{x}_{\rm tot}}
P_{\delta-\alpha}(k) \right]+ \nonumber \\ & & \left[ \beta^2
P_{\delta}(k) + \left( \frac{x_{\alpha}} {\tilde{x}_{\rm tot}}
\right)^2 P_{\alpha}(k) + \frac{2 \beta x_{\alpha}} {\tilde{x}_{\rm
tot}} P_{\delta-\alpha}(k) \right]\ , \label{powTb} \eeqa where
$\tilde{x}_{\rm tot} \equiv (1 + x_{\rm tot}) x_{\rm tot}$,
$P_{\delta}$ and $P_{\alpha}$ are the power spectra of the
fluctuations in density and in $x_{\alpha}$, respectively, and the
power spectrum $P_{\delta-\alpha}$ is the Fourier transform of the
cross-correlation function,
\beq \xi_{\delta-\alpha} (r) = \langle \delta(\br_1)\, 
\delta_{\rm x_{\alpha}} (\br_1 + \br) \rangle\ . \label{xi} \eeq 
Here $\beta$ is given by \citep{BL04} \beq \beta = 1 + \frac{x_c}
{\tilde{x}_{\rm tot}} + (\gamma_a - 1) \left[ \frac{T_{\gamma}} {T_k -
T_{\gamma}} + \frac{x_c} {\tilde{x}_{ \rm tot}}\, \frac{d
\log(\kappa_{1-0})} {d \log(T_k)} \right] \ ,\eeq
where the adiabatic index is $\gamma_a = 1 + (\delta_{T_k} / \delta)$.
Also, $\mu$ is the cosine of the angle between the wavenumber $\bk$ of
the Fourier mode and the line-of-sight direction. Considering
equation~(\ref{powTb}) as a polynomial in $\mu$, i.e.,
\begin{equation}
P_{T_b}(\bk) = \mu^4 P_{\mu^4} + \mu^2 P_{\mu^2} + P_{\mu^0}\ , 
\end{equation}
we note that the power at just three values of $\mu$ is required in
order to separate out observationally the coefficients of 1, $\mu^2$,
and $\mu^4$ for each $k$.  In particular, we can construct a
combination that probes whether some sources of flux fluctuations are
uncorrelated with $\delta$, through \citep{BL04}
\begin{equation}
P_{\rm un-\delta}(k) \equiv P_{\mu^0}- {P_{\mu^2}^2\over 4 P_{\mu^4}}=
\left( \frac{x_{\alpha}} {\tilde{x}_{\rm tot}} \right)^2\, 
\left(P_{\alpha} - {P_{\delta-\alpha}^2 \over P_{\delta}}\right)\ . 
\label{eq:un-d} \end{equation}
Note that since 21cm measurements on large angular scales cannot yield
significant information \citep{BL04}, we focus in this paper on
three-dimensional power spectra, which are directly measurable from
21cm observations in small angular fields.

We note that several additional redshift effects do not significantly
modify the 21cm brightness fluctuations. First, consider an observed
correlation function (or the equivalent power spectrum) that is
calculated by averaging the product of the 21cm temperatures at pairs
of points within a small three-dimensional volume. We assume that this
volume is centered on redshift $z$ and has a redshift width $\Delta z$
that corresponds to the observed frequency range used to calculate the
correlation function at $z$. The change in the cosmic mean temperature
$T_b(z')$ with redshift $z'$ within this slice would produce
variations in temperature even in an homogeneous expanding
universe. However, these variations are coherent on the sky and have
a very simple form. Denoting by $l$ the line-of-sight component of
$\br$, where $l=0$ marks the center of the redshift slice and
$\Delta l$ its width, the correlation function $\xi_{T_b} (r)
\equiv \langle \delta_{T_b} (\br_1) \delta_{T_b} (\br_1 + \br) \rangle
$ is modified to $\langle (\delta_{T_b} (\br_1) + l_1\, d_l T_b
)\, [\delta_{T_b} (\br_1 + \br) + (l_1 + l)\, d_l T_b ]
\rangle$, where we denote the derivative $[dT_b(z)/d l]$ at $z$ by
$d_l T_b$. Noting that $\langle \delta_{T_b} (\br_1) \rangle$ and
$\langle l_1 \rangle$ are both zero within the slice, we find that
$\xi_{T_b} (r)$ receives an extra term, $(d_l T_b)^2 \langle
l_1^2 \rangle$. Since $\langle l_1^2 \rangle = (\Delta
l)^2/12$, the finite redshift width $\Delta z$ adds to the power
spectrum $P_{T_b}(\bk)$ a term
\beq P_{T_b:\, \Delta z} (\bk) = \frac {2} {3} \pi^3 \left[ \Delta z
\frac {dT_b(z)} {dz} \right]^2 \delta_D^3 (\bk)\ . \eeq Thus,
this effect does not modify the power spectrum at $k \ne 0$.

In addition, peculiar velocities change the apparent distance along the
line of sight. Over a distance $l$, the unperturbed velocity $v=a H
l$ receives a fractional perturbation $\delta_v(l)$ which is of order
the density perturbation $\delta$ averaged on the scale $l$ (e.g.,
$\delta_v= -\delta/3$ for a spherical tophat perturbation out to
radius $l$). The mean temperature $T_b$ is modified due to this
displacement by $\sim l\, \delta_v(l)\, d_l T_b$, which is
smaller than $\delta$ by $\sim a l H /c$ which at $z=20$ is $ l/(2\,
{\rm Gpc})\ll 1$. Also, to first order we can neglect the effect of
the peculiar velocities on the perturbation $\delta_{T_b}$
itself. Finally, gravitational potential fluctuations are $\sim
10^{-5}$ on the scale of the horizon and are suppressed relative to
the density fluctuations by a factor of $\sim (a l H /c)^2$ on small
scales. Note that $10^{-5}$ fluctuations are important in CMB
observations, since at $z \sim 1100$ density fluctuations on small
scales had not yet grown due to the radiation pressure of the
photon-baryon fluid; in the redshift range that we consider for 21cm
observations, however, density fluctuations are already at a level of
$\sim 1\%$ on 100 Mpc scales (at $z=20$), and it would be very
difficult to measure the $\sim 10^{-4}$ fluctuations on Gpc scales
\citep{BL04}.

\section{The \Lya Flux of Galaxies}

\label{sec:flux}

The intensity of \Lya photons observed at a given redshift is due to
photons that were originally emitted between the rest-frame
wavelengths of \Lya and the Lyman limit. Photons that were emitted
below \Lyb by some source simply redshift until they are seen by an
atom at $z$ at the \Lya wavelength. Such photons can only be seen out
to a distance corresponding to the redshift $z_{\rm max}(2)$, where
$1+z_{\rm max}(2) = (32/27)\times (1+z)$, where $32/27$ is the ratio
of \Lya to \Lyb wavelengths. Photons above the \Lyb energy redshift
until they reach the nearest atomic level $n$. The neutral
intergalactic medium (IGM) is opaque even to the higher levels and so
the photon is absorbed and re-emitted. During these scatterings, the
photon is almost immediately downgraded to a
\Lya photon and then keeps on scattering, and so most of the atoms
that interact with it encounter it as a \Lya photon. To be seen at the
\Lya resonance at $z$, the photon must have been emitted below a
redshift given by \begin{equation} 1+z_{\rm max}(n) = (1+z)
{\left[1-(n+1)^{-2}\right] \over(1-n^{-2})} \ . \end{equation} Thus,
we simply add up the number flux of photons emitted between
consecutive atomic levels, and integrate over sources out to the
distance corresponding to the appropriate redshift interval.

Note that while sources are expected to also produce a strong \Lya
line, due to internal reprocessing of ionizing photons into \Lya
photons, these photons are not expected to have a significant
effect. The photons stream into the \ion{H}{2} region of their source,
and by the time they reach the edge, many of them have redshifted
above the \Lya wavelength by more than the line width; the remaining
ones are trapped and diffuse only across a distance of $\sim $ tens of
parsecs before they too redshift out of the line. Thus, these photons
only affect a very small fraction of the IGM, much less than the
ionized fraction, which is itself small in the redshift range of
interest here. Even though the \Lya flux within this volume can be
high, the brightness temperature saturates when $x_{\rm tot}
\rightarrow \infty$. Observations on scales larger than the size of 
an \ion{H}{2} region (which is $\la 1$ comoving Mpc) smooth the signal
over many such regions. The overall correction is thus small both on
the mean brightness temperature and on the fluctuations in it.

We therefore get \citep{Madau} \beq x_{\alpha} = \frac {16 \pi^2
T_\star e^2 f_{\alpha}} {27 A_{10} T_{\gamma} m_e c} S_{\alpha}
J_{\alpha}\ , \label{eq:xa} \eeq
where $J_{\alpha}$ is the proper \Lya photon intensity (defined as the
spherical average of the number of photons hitting a gas element per
unit area per unit time per unit frequency per steradian), and
$S_{\alpha}$ \citep{Miralda} is a correction factor between 1 and 2
that depends on the gas temperature and accounts for the
redistribution of photon frequencies near the \Lya line center due to
scattering off the thermal distribution of atoms. The intensity is
itself given by a sum over $n$ (see also \S~\ref{sec:Pk}), \beq
J_{\alpha}=\frac{(1+z)^2} {4 \pi} \sum_{n=2}^{n_{\rm max}} 
\int_z^{z_{\rm max}(n)} \frac{c dz'}{H(z')} \epsilon(\nu_n', z')\ , 
\label{eq:Ja} \eeq where absorption at level $n$ at redshift $z$ 
corresponds to an emitted frequency at $z'$ of
\beq \nu_n' = \nu_{\rm LL} (1-n^{-2}) {(1+z')\over (1+z)}\ , 
\label{eq:nu} \eeq in terms of the Lyman limit frequency 
$\nu_{\rm LL}$. The comoving photon emissivity (defined as the number
of photons emitted per unit comoving volume, per proper time and
frequency, at rest-frame frequency $\nu$ at redshift $z$) is in turn
\beq \epsilon(\nu, z) = \bar{n}_b^0\, f_* \frac{d}{dt} F_{\rm gal}
(z)\ \epsilon_b(\nu)\ , \label{eq:emit} \eeq where $\bar{n}_b^0$ is
the cosmic mean baryon number density at $z=0$, $f_*$ is the
efficiency with which gas is turned to stars in galactic halos,
$F_{\rm gal}(z)$ is the gas fraction in galaxies at $z$, and
$\epsilon_b(\nu)$ is the spectral distribution function of the sources
(defined as the number of photons per unit frequency emitted at $\nu$
per baryon in stars).

To determine the gas fraction in galaxies, we must first find the halo
mass function. Although the basic model of \citet{ps74} describes the
results of numerical simulations qualitatively, we use the halo mass
function of \citet{shetht99} which fits numerical simulations more
accurately \citep{jenkins}. In order to calculate fluctuations among
large-scale regions of various densities in the global gas fraction
that resides in galaxies within them, we must also find the halo mass
function as a function of the overall mean density in each region. We
adjust the mean halo distribution as a function of density based on
the prescription of \citet{Bar04a}, where we showed that this
prescription fits a broad range of simulations including ones run at
$z \sim 20$. As gas falls into a dark matter halo, it can fragment
into stars only if its virial temperature is above $10^4$K for cooling
mediated by atomic transitions, or $\sim 500$ K for molecular ${\rm
H}_2$ cooling.  Given a minimum cooling mass, we consider two
possibilities for the star formation efficiency: either a fixed
efficiency in all halos, as in equation~(\ref{eq:emit}), or an
efficiency that drops as the square of the halo circular velocity
$V_c$ at $V_c \la 180~{\rm km~s^{-1}}$, as suggested by local
observations \citep{Dekel,Kauffman}; in the latter case, we integrate
over the halo mass function in order to determine the net effective
value of $f_*$.

The spectral distribution function $\epsilon_b(\nu)$ depends on the
assumed sources of radiation. In our discussion, we assume that stars
dominate over mini-quasars, as expected for the low-mass galaxies that
are typical at high redshift \citep{WL04}. The stellar spectrum
depends on the initial mass function (IMF) and the metallicity of the
stars. In what follows, we consider two examples. The first is the
locally-measured IMF of \citet{scalo} with a metallicity of $1/20$ of
the solar value, which we refer to as Pop II stars. The second case,
labeled as an extreme Pop III IMF, consists entirely of zero
metallicity $M \ga 100\, M_{\odot}$ stars, as expected for the
earliest generation of galaxies \citep{Bromm04}. In each case, we
approximate the emissivity as a separate power law $\epsilon_b(\nu)
\propto \nu^{\alpha_S - 1}$ between every pair of consecutive levels
of atomic hydrogen. For example, using \citet{Leith99} we find that
the Pop II stars emit 9,690 photons per baryon between \Lya and the
Lyman limit, out of which 6,520 photons are between \Lya and \Lyb with
an $\alpha_S = 0.14$ in this lowest frequency interval. The
corresponding numbers for Pop III stars [using \citet{Bromm}] are
4,800 , 2,670 , and $\alpha_S = 1.29$.

\begin{figure}
\plotone{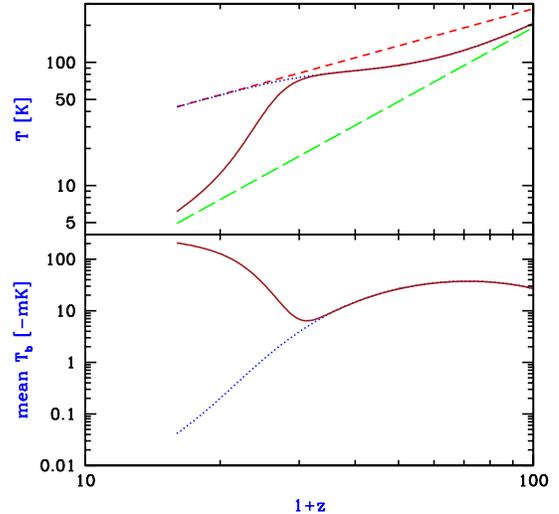}
\caption{Redshift evolution of various mean temperatures.
The mean 21cm spin temperature is shown in the upper panel for
adiabatic IGM gas with \Lya coupling (solid curve) and without it
(dotted curve). Also shown for comparison are the gas temperature
(long-dashed curve) and the CMB temperature (short-dashed curve). The
mean 21cm brightness temperature offset relative to the CMB is shown
in the lower panel, with \Lya coupling (solid curve) and without it
(dotted curve). The \Lya flux is calculated assuming that galaxies
with Pop III stars form in dark matter halos where the gas cools
efficiently via atomic cooling. The star formation efficiency is
normalized so that the \Lya coupling transition (i.e., $x_{\rm
tot}=1$) occurs at redshift 20.}
\label{Tevol}
\end{figure}

Figure~\ref{Tevol} illustrates the cosmic evolution of the various key
temperatures. The residual electron fraction keeps the gas temperature
$T_k$ locked to the CMB temperature [$T_\gamma = 2.725 (1+z)$ K] until
$1+z\approx 142 (\Omega_b h^2/0.024)^{2/5}$ \citep{Peebles}, but
subsequently the gas cools adiabatically. The mean temperature offset
$T_b$ is $\sim -30$ mK at $z=100$ and it rises towards zero at low
redshift until the \Lya flux from galaxies drops it again, to -100 mK
and below. Note that this Figure and our results in subsequent
sections assume an adiabatically cooling cosmic gas, except where we
explicitly assume pre-heating by X-rays.

\section{Fluctuations in \Lya Flux}

\label{sec:basic}

There are two separate sources of fluctuations in the \Lya flux. The
first is density inhomogeneities.  Since gravitational instability
proceeds faster in overdense regions, the biased distribution of rare
galactic halos fluctuates much more than the global dark matter
density \citep{k84, Bar04a}. When the number of sources seen by each
atom is relatively small, Poisson fluctuations provide a second source
of fluctuations, statistically independent of the first source, to
linear order. Unlike typical Poisson noise, these fluctuations are
correlated between gas elements at different places, since two nearby
elements see many of the same sources, where each source is at about
the same distance from the two gas elements.

Since each hydrogen atom receives some \Lya flux from sources as far
away as $\sim 250$ comoving Mpc, the flux is naively expected to be
highly uniform based on the homogeneity of the universe on these large
scales. However, the fluctuations in flux are actually relatively
large because a significant portion of the flux comes from nearby
sources. One reason is the $1/r^2$ decline of flux with distance. In
an homogeneous, non-expanding universe, the flux per $\log(r)$ would
scale linearly with $r$, since the $1/r^2$ decline of the flux is
overcome by the $r^3$ volume factor; while large scales would dominate
the total flux, the relative contribution of small scales would
already be far more significant than if we were simply counting galaxy
numbers, where the overall scaling would be cubic instead of linear in
$r$. The $1/r^2$ scaling of flux also strongly magnifies Poisson noise
since, e.g., the 21cm emission along a given line of sight can
fluctuate strongly due to a single \Lya source that happens to lie
very close to that line of sight. Although the mean flux at a
point is well defined, if we calculate the mean squared flux at a
point, then a shell at a distance $r$ contains a number of sources
$\propto r^2$ and contributes to the squared flux a value $\propto
1/r^4$; thus, the mean squared flux diverges at small radii, and since
the correlation function of flux at $r=0$ is proportional to its mean
square, it too diverges at small $r$.

A second reason for the dominance of small scales in the \Lya flux is
that higher Lyman series photons, which are degraded to \Lya photons
through scattering as discussed in the previous section, can only be
seen from a small redshift interval that corresponds to the wavelength
interval between two consecutive atomic levels of hydrogen. A third
reason is that distant sources are observed after a time delay
determined by the speed of light, and their relative contribution is
reduced since galaxy formation was not as advanced at earlier
times. To facilitate the weighting of different scales, we summarize
equations~(\ref{eq:xa}), (\ref{eq:Ja}), and (\ref{eq:emit}) in the
form: \beq x_{\alpha} =\int_{z'=z}^{z_{\rm max}(2)} \frac{d
x_{\alpha}} {d r} \frac{dr}{dz'} dz'\ . \label{eq:flux} \eeq

\begin{figure}
\plotone{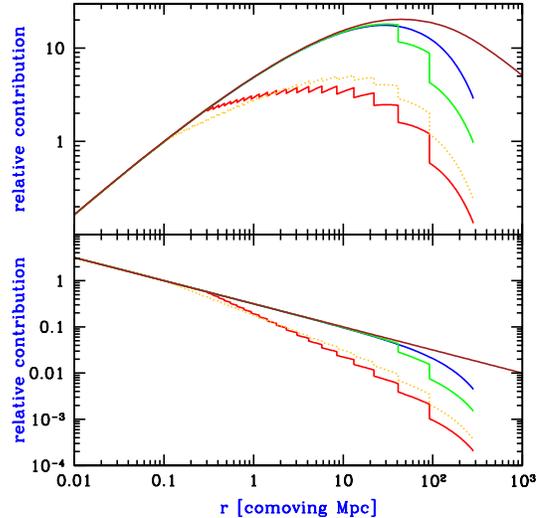}
\caption{Relative contribution per $\log(r)$ of shells at various radii 
$r$ from a gas element at redshift $z=20$, to the fluctuations in the
\Lya flux. We consider the fluctuations sourced by density
inhomogeneities (top panel) and by Poisson fluctuations (bottom
panel). In each set of solid curves, we show from top to bottom (at
large $r$) the asymptotic case of a non-expanding universe followed by
galaxies, where we include stellar radiation emitted up to Ly$\beta$,
Ly$\delta$, or the case of full Lyman-band emission (see text). The
minimum galaxy mass is set by the atomic cooling threshold in dark
matter halos. Inside galaxies, we assume a fixed star formation
efficiency and a Pop III stellar population. Also shown for comparison
is the case of full Lyman-band emission with Pop II stars (dotted line). All
curves are arbitrarily normalized to unity at $r=0.1$ Mpc.}
\label{radialfig}
\end{figure}

The top panel of Figure~\ref{radialfig} shows an estimate of the
relative contribution to the fluctuations in \Lya flux from sources at
various distances from a gas element at $z=20$. We take the
contribution to the flux per $\log(r)$, i.e., $d x_{\alpha}/d\log r$
(which is $\propto r$ at small distances), and multiply it by the mean
density fluctuation in spheres of radius $r$, in order to estimate the
contribution per $\log(r)$ to the density-induced fluctuations in
flux. We assume a scale-invariant spectrum of primordial density
fluctuations, as predicted by inflation and indicated by the latest
observations \citep{Sel04}. Since this spectrum yields fluctuations
that diverge only logarithmically at small scales, the declining
contribution of small scales to the flux dominates as $r
\rightarrow 0$. On the other hand, density fluctuations decrease as
$r^{-2}$ on scales $r \ga 100$ Mpc, which overcomes the $\propto r$
contribution to the flux and reduces the role of large-scale modes in
the flux fluctuations. Although these effects make the $\sim 10$--100
Mpc scales dominate the density-induced fluctuations in \Lya flux, the
drop in the density of time-retarded sources (as evident from the
comparison of the top two curves in the top panel), plus the large
contribution of nearby sources through photons above \Lyb, make the
overall effect smoothed out over scales of $\sim 0.1$--100 Mpc. The
figure shows discontinuous jumps at distances that correspond to the
various redshifts $z_{\rm max}(n)$ that mark the \Lya horizons for
photons emitted with energies between the $n \rightarrow 1$ and $n+1
\rightarrow 1$ transitions of hydrogen.

The bottom panel of Figure~\ref{radialfig} shows the relative
contribution to Poisson-induced fluctuations in the \Lya flux. We
construct this estimate by multiplying $d x_{\alpha}/d\log r$ by the
Poisson $1/\sqrt{N}$ fluctuation, where for small distances $N \propto
r^3$.  In a non-expanding universe, this would already diverge as
$1/\sqrt{r}$ at small scales. The previously-mentioned effects only
increase the dominance of small scales in the Poisson
fluctuations. Thus, measuring the two different sources of flux
fluctuations has a complementary effect and allows a broad range of
scales to be probed. Furthermore, the Poisson fluctuations are
independent (to linear order) of the underlying primordial density
fluctuations, and more directly measure the number density of
sources. In Figure~\ref{radialfig} and the following figures, we
include in the case of ``full Lyman-band emission'' all the photons
emitted from \Lya up to a maximum range given by the interval between
levels $n_{\rm max}$ and $n_{\rm max}+1$ of hydrogen (assuming that
none of the photons are obscured or absorbed by dust within the host
galaxy). We set this maximum level by cutting off photons that could
only be seen from sources so nearby that the gas element would have to
be within the \ion{H}{2} region of the individual source, and thus
this gas would be ionized and would not contribute to the 21cm optical
depth. Calculating a typical \ion{H}{2} region size based on an escape
fraction of ionizing photons of $f_{\rm esc}=0.3$, this yields a
maximum level $n_{\rm max}=23$ in the Pop III case at $z=20$, and
corresponds to including $99.3\%$ of all photons emitted between \Lya
and the Lyman limit. We note that since ionizing sources are clustered
\citep{WL04, Bar04a, Fur04b}, the appropriate cutoff scale might be
somewhat larger.

\section{Power Spectrum of Fluctuations in the 21cm Flux}

\label{sec:Pk}

We first consider the effect of density perturbations on the flux as
given by equation~(\ref{eq:flux}). When considering perturbations, we
let $r$ be the Lagrangian position, which corresponds to the comoving
distance at some initial high redshift. We now consider a field of
perturbations $\delta_0(\br)$, linearly extrapolated to the present
from the initial time with the growth factor $D(z')$, so that the
perturbations at a redshift $z'$ are $\delta_0(\br) D(z')$. Consider
the contribution of a Lagrangian volume $d V = dA dr$ at position
$\br$ to the flux observed at the origin of the coordinates at
redshift $z$. The emission rate $\epsilon(\nu', z')$ is different than
average, since the density perturbation $\delta_0(\br) D(z')$ changes
the abundance of galaxies. On large scales, the change in the
abundance of halos is linear in the density perturbation, for every
halo mass \citep{mw96, Bar04a}, yielding a weighted bias $b(z')$ in
the overall emission rate.  Thus, the emission rate is increased by a
factor $[1 + b(z') \delta_0(\br) D(z')]$. Note that the required
emitted frequency is determined by the redshift just as in the
unperturbed case [i.e., equation~(\ref{eq:nu})], and also to first
order we can use the unperturbed $z'(r)$ relation in the perturbation
term because corrections to this are small. The observed proper
intensity at $z$ due to a proper emitting area $dA_p$ at the source at
redshift $z'$ is unchanged by perturbations, due to the conservation
of surface brightness known from gravitational lensing
theory. However, for a fixed Lagrangian area $dA$, the proper area
$dA_p$ (and thus also the observed intensity) is increased by the
factor $[1+ \frac{2}{3} \delta_0(\br) D(z')]$ to linear
order. Finally, we must relate the remaining factor in the volume
element $dV'$, i.e., the radial width $dr$, to the redshift interval
$d z'$. The appropriate relation is $dz' = [dv_r/d r] d r/c$, where
$v_r$ is the radial component of the physical velocity. The
unperturbed form of this Doppler relation is $dz' = H(z') dr/[(1+z')
c]$, which accounts for the denominator in equation~(\ref{eq:Ja}).
Even in the perturbed case, the radial integration limit is fixed in
redshift as in equation~(\ref{eq:Ja}). Note that we have assumed that
the retarded time corresponding to a given source redshift $z' > z$ is
unperturbed; the perturbation in peculiar velocity causes a $\Delta z
\sim \delta a r/(c/H)$, which is negligible compared to the biased
galaxy fluctuation which corresponds to a $\Delta z \sim \delta (1+z)$
\citep{Bar04a}.

Integrating over the volume $dV'$, we add the various terms mentioned
before and obtain the total linear perturbation in $x_{\alpha}$, which
is a spatial average of the density perturbation field, plus a spatial
average of the velocity gradient term. The fractional perturbation in
$dv_r/d r$ has a simple Fourier transform, $\td_{d_rv_r} = -\mu^2 \td$
\citep{kaiser, Indian}, which is used to separate the angular structure 
of the power-spectrum [see \S~\ref{sec:sep} and \citet{BL04}]. Here,
$\mu$ is the cosine of the angle between the wavenumber $\bk$ of the
Fourier mode and the $\br$ (i.e., line of sight) direction. After
proper averaging we find the Fourier perturbation in $x_{\alpha}$ in
terms of the Fourier transform of the density perturbation at $z$:
\beq
\td_{\alpha} (\bk) =
\tilde{W}(k) \td (\bk)\ ,\eeq where
\beq \tilde{W}(k) = \frac{1}{x_{\alpha}} \int_z^{z_{\rm max}(2) } dz' 
\frac{d x_{\alpha}} {d z'} \frac {D(z')} {D(z)} \left\{ [1 + b(z')] 
j_0(k r) - \frac{2}{3} j_2(k r) \right\}\ ,
\label{eq:Wk} \eeq expressed in terms of spherical Bessel functions. 
The first term is reminiscent of the effective bias in galaxy number
density, where the addition of $1$ to the Lagrangian bias accounts for
the higher physical density per unit volume. The second term is a
geometrical correction for the case of flux, which arises from the
velocity gradient term. The contribution to the power spectrum
$P_{\alpha}(k)$ is then $\tilde{W}^2(k) P_{\delta}(k)$, while the
contribution to $P_{\delta-\alpha}(k)$ is $\tilde{W}(k)
P_{\delta}(k)$. We note that $P_{\alpha}$ corresponds to a real-space
correlation function of the flux at two points $A$ and $B$ separated
by a comoving distance $l$. In this correlation, if point $A$ is
closer to sources at some location $\br$ then point $B$ will see the
progenitors of the galaxies seen by point $A$. The relation between
the flux of the progenitors and the flux of the later parent is
implicitly accounted for in the above weighting by the
redshift-dependent bias factor $b(z')$.

If the number of galaxies $N$ that contribute to the flux is
relatively small, then in addition to fluctuations in the mean number
density and mass distribution of halos we expect significant Poisson
fluctuations. We first derive a general framework for analyzing
correlations induced by Poisson fluctuations. We note that in the
context of the ionizing intensity due to quasars, \citet{z92} derived
the joint probability distribution of the intensities at two different
points. Focusing on the two-point correlation function only, we
present here a derivation that is simpler and more general, including
cosmological redshift and time-retardation effects.

We consider a large number of small volumes $dV$, such that the mean
number of halos $\langle dn \rangle \ll 1$ in each, so that $dn^2=dn$
since at most a single halo may be found within a single volume
element. Although we describe a volume distribution for convenience,
this setup also applies to a generalized notion of volume, where the
population of halos is divided up into many small units, and the
effective position within this ``volume'' is specified by as many
parameters as needed, in addition to spatial position (e.g., halo
mass, emission time, merger history, etc.). We denote by
$dx_{\alpha}^A$ the flux that arrives at a point $A$ due to $dn$ halos
in some volume element; this flux is proportional to $dn$, with a
proportionality factor that may depend on the distance to $A$ and on
the various parameters that specify the volume element, and can
incorporate all appropriate redshift effects. Then the mean expected
flux $x_{\alpha}$ at a point $A$ is $\sum [dx_{\alpha}^A/dn]
\langle dn \rangle$, while the expectation value of the flux at point 
$A$ times the flux at point $B$ is $\sum \sum [dx_{\alpha}^A/dn_1]
[dx_{\alpha}^B/dn_2] \langle dn_1 dn_2 \rangle$. When the number of
small volume elements is large, the double sum over different halos
$dn_1 \ne dn_2$ yields an expectation value $x_{\alpha}^A
x_{\alpha}^B$ which is simply $x_{\alpha}^2$, while the sum over $dn_1
= dn_2$ (which we then denote simply $dn$) yields the correlated
Poisson term. Thus, the correlation function of relative Poisson
fluctuations in flux is \beq \xi_P(l) = \frac{1} {x_{\alpha}^2} \sum
\frac{dx_{\alpha}^A} {dn} \frac{dx_{\alpha}^B} {dn} \langle dn \rangle 
\ , \label{eq:xiP} \eeq where the points $A$ and $B$ are separated by 
a distance $l$. 

To apply this equation, we divide halos into effective
four-dimensional volumes $dVdM$, which include three spatial
dimensions as well as a distribution in halo mass, and convert the sum
into a four-dimensional integral. In a homogeneous, non-expanding
universe, the only difference between $dx_{\alpha}^A /dn$ and
$dx_{\alpha}^B /dn$ would be the distance factors $1/r_A^2$ and
$1/r_B^2$. For distant sources, $r_A \sim r_B \gg l$, we have an
integral $dV/r^4$ at all $r \ga l$, so the contribution to $\xi_P(l)$
of all such sources is $\propto 1/l$. On the other hand, sources near
$A$ satisfy $r_A \ll l$ and $r_B \approx l$, so their contribution is
$dV/(r_A^2 l^2)$, over $r_A$ from 0 to $\sim l$, which is again
$\propto 1/l$. Sources near $B$ contribute equally, and the overall
correlation function diverges at small separations as $\xi_P(l)
\propto 1/l$, as expected from the simple scaling argument at the end
of \S~\ref{sec:basic}.

To apply this formalism in the present context, we first write the
flux as a double integral over the the halo number density as a
function of mass and volume. Halos in a comoving volume $dV$ at $\br$
contribute a flux $\propto 1/r^2$ at the origin, and we assume that
the fraction of the flux contributed by halos of a given mass $M$ is
proportional to the gas fraction contained in these halos. Thus, we
write the contribution $dx_{\alpha}$ from the halos in $dV$ as \beq
dx_{\alpha} \equiv P(z')\,\frac{1}{r^2}\, \int_M M \frac{dn(z')}{dM}
dM dV\ , \label{eq:Pz} \eeq where we have factored out explicitly the
$1/r^2$ dependence. Using the \citet{shetht99} halo mass function
$dn/dM$ at $z'$, and $dx_{\alpha}$ calculated as in \S~\ref{sec:flux},
the above equation defines $P(z')$ as an overall normalization factor
that accounts for the spectral distribution and redshift effects. Note
that the mean intensity $x_{\alpha}$ at a point is given by the volume
integral of equation~(\ref{eq:Pz}), which yields the same result as
given by equation~(\ref{eq:flux}).

An additional subtlety necessitates a small adjustment in
equation~(\ref{eq:xiP}) before we can apply it in the present
context. This equation, as written, requires that the same halos be
seen by points $A$ and $B$ at the same redshift $z$. However, if,
e.g., point $A$ is closer to some location $\br$, then the point $B$
will see the progenitors of the galaxies at $\br$ seen by point
$A$. When we apply equation~(\ref{eq:xiP}), we integrate over the halo
distribution seen by the closer point $A$, and note that the total
emission rate of its progenitors at the higher redshift $z'_B \equiv
z'(r_B)$ is on average some fraction of the emission rate of the later
parent galaxy at $z'_A \equiv z'(r_A)$. An actual correlation function
measured over a significantly large volume will indeed average over
the progenitor distribution. This average fraction may depend on
parent mass and on the two redshifts. First, we neglect the dependence
on parent mass, i.e., we consider a mass-averaged fraction. This is
likely to yield an accurate result at high redshift, since when
galaxies are rare they are all found in a narrow range of halo masses
just above the threshold mass for star formation. We then fix the
redshift dependence of this factor so that the total gas fraction in
progenitors at $z'_B$ is consistent with the total fraction calculated
in the normal way from the halo mass function at $z'_A$. Thus the
dimensionless correlation function is
\beq \xi_P(l) = \frac{2} {x_{\alpha}^2} \int_V dV \int_M
\frac{dn(z'_A)}{dM} dM\, M^2 \frac{P(z'_A)}{r_A^2}
\frac{P(z'_B)} {r_B^2} \frac{F_{\rm gal}(z'_B)} {F_{\rm gal}(z'_A)} \
,\eeq where $dV=d^3r_A$, and we integrate over half the volume where
$r_A < r_B$, with the factor of 2 accounting for the contribution of
sources that are closer to $B$.

We have assumed in calculating $r_B$ that the progenitors of a galaxy
are at the same location as the parent. The actual displacements are
at most of order the Lagrangian size of a typical galactic halo, $\sim
0.1$ comoving Mpc, which is negligible when $r_B \gg 0.1$ Mpc, while
if we consider small values of $r_B$ then the displacements resulting
from gravitational infall velocities ($\la c/1000$) during the time
interval $|r_B - r_A|/c$ are still a small fraction of the distance
$r_B -r_A$.

\begin{figure}
\plotone{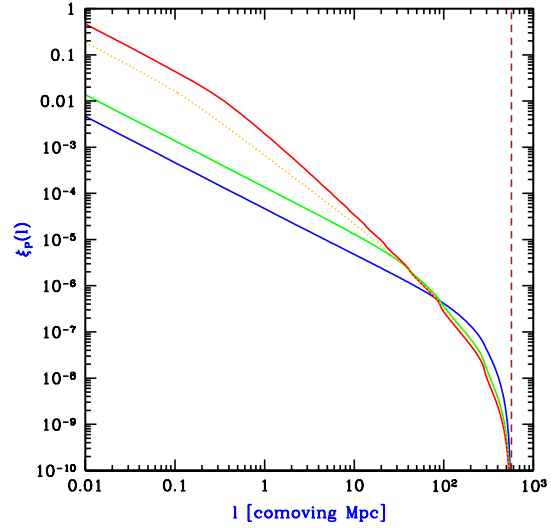}
\caption{The Poisson-induced dimensionless correlation function of 21cm 
brightness fluctuations versus comoving separation. We assume that
galaxies formed via atomic cooling in halos at $z=20$ with a fixed
star formation efficiency in all halos, set to produce the coupling
transition at this redshift (so that $x_{\rm tot}=1$). We also assume
that the IGM gas has been cooling adiabatically down to this
redshift. In each set of solid curves we include, from bottom to top
at $l=1$ Mpc, stellar radiation emitted up to Ly$\beta$, Ly$\delta$,
or full Lyman-band emission (see \S~\ref{sec:basic}), assuming Pop III
stars. Also shown for comparison is the case of full Lyman-band
emission with Pop II stars (dotted line). We also show the separation
equal to twice the maximum Ly$\alpha$ horizon (dashed line).}
\label{Xifig}
\end{figure}

The correlation function $\xi_P(l)$ drops sharply at large distances,
and in particular, when $l$ is more than twice the distance
corresponding to $z_{\rm max}(2)$, no halo can contribute flux to both
points $A$ and $B$ and thus $\xi_P(l)=0$. At small distances, $\xi_P(l)
\propto 1/l$ as in a non-expanding universe, but even for $l \rightarrow 0$ 
the proportionality factor is affected by distant sources since they
dominate the flux [compare \citet{z92}]. Figure~\ref{Xifig}
illustrates this correlation function in several cases. The maximum
Ly$\alpha$ horizon in this case, corresponding to the redshift $z_{\rm
max}(2)$, is 284 comoving Mpc. When we include only the stellar
radiation emitted up to Ly$\beta$, $\xi_P(l)$ begins to drop below the
$1/l$ asymptotic form at separations $l \ga 200$ Mpc and sharply drops
to zero at twice the Ly$\alpha$ horizon. However, when more energetic
radiation is included, $\xi_P(l)$ drops below the $1/l$ power law at
much smaller $l$, corresponding to the smaller effective source
horizons of the energetic photons, thus making the cutoff at $l \sim
200$ Mpc much rounder and less prominent.

In this Figure, the star formation parameters are fixed to produce the
\Lya coupling transition (i.e., $x_{\rm tot}=1$) at $z=20$. In our
standard case of full Lyman-band emission and Pop III stars, with only
atomic cooling occurring in galactic halos, the required star formation
efficiency is relatively low, $0.22\%$, and $f_{\rm esc}=0.3$ yields
an ionized fraction of $1\%$ at $z=20$, with complete reionization at
$z=9$. Although here we have calculated $\xi_P(l)$, in the rest of the
paper we Fourier transform the Poisson-induced correlation function
and find the corresponding power spectrum.

For the case of a mass-dependent star formation efficiency, we insert
the appropriate weighting factor into the various relations given in
this section. Note that if star formation in individual galaxies
actually proceeded in short, random bursts with a duty cycle $\eta \ll
1$, then the higher rate of star formation during bursts (by a factor
$1/\eta$) would counterbalance the fraction of time of burst activity
($\eta$), and the overall correlation function would be unchanged for
both of the sources that we consider for fluctuations in flux.

We insert two additional cutoffs that are expected on small
scales. The first is the cutoff due to baryonic pressure, which
affects the density fluctuations and has an approximate shape $(1+ k^2
R_{\rm F}^2)^{-2}$ \citep{sgb94, gh98} where the filtering scale is
$R_{\rm F} = 2.9 \left({M_{\rm F}}/{10^6 M_{\odot}} \right)^ {1/3}
\left({\Omega_m h^2/0.14}\right)^{-1/3} {\rm kpc}$, in terms of the
filtering mass $M_{\rm F}$ which is a weighted time average of the
Jeans mass\citep{gh98}. We calculate $M_{\rm F}$ appropriately for the
two cases we consider, an adiabatically cooling IGM or a pre-heated
IGM. The second cutoff that we include results from the thermal width
of the 21cm line which also smooths out the fluctuations on small
scales. We approximate this effect as a Gaussian cutoff, $e^{-k^2
R_{\rm T}^2}$, where the spatial scale corresponding to the
one-dimensional thermal velocity distribution is
\beq R_{\rm T} = 7.0 \left( \frac{m_p}{1.22} \right)^{-1/2}
\left( \frac{T_k}{100\, {\rm K}} \right)^{1/2} \left( \frac{1+z}{10} 
\right)^{-1/2} \left(\frac {\Omega_m h^2} {0.14}\right)^{-1/2} 
{\rm kpc}\ , \eeq and $m_p$ is the mean mass per particle of neutral
primordial gas in units of the proton mass.

\begin{figure}
\plotone{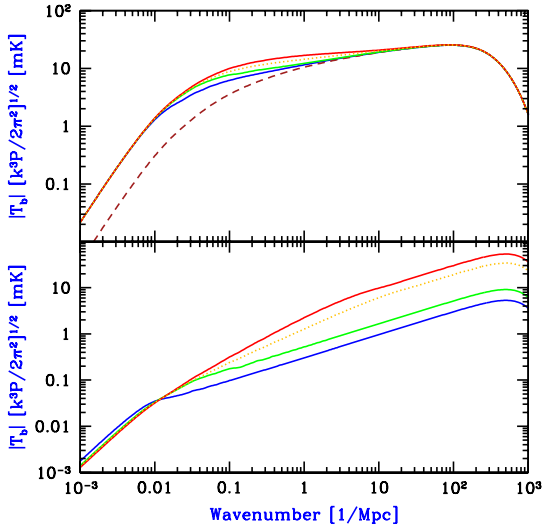}
\caption{Power spectra of 21cm brightness fluctuations versus comoving 
wavenumber. Shown are $P_{\mu^2}$ (top panel), which contains the
contribution of the density-induced fluctuations in flux, and $P_{\rm
un-\delta}$ (bottom panel), which is solely due to the Poisson-induced
fluctuations. We assume that galaxies formed via atomic cooling in
halos at $z=20$ with a fixed star formation efficiency in all halos,
set to produce the coupling transition at this redshift (so that
$x_{\rm tot}=1$). We also assume that the IGM gas has been cooling
adiabatically down to this redshift. In each set of solid curves we
include, from bottom to top at $k=0.1$ Mpc$^{-1}$, stellar radiation
emitted up to Ly$\beta$, Ly$\delta$, or full Lyman-band emission (see
\S~\ref{sec:basic}), assuming Pop III stars. Also shown for comparison
is the case of full Lyman-band emission with Pop II stars (dotted
line). We also show $2 \beta P_{\delta}$ (dashed curve).}
\label{Pkfig}
\end{figure}

For the same parameters as in Figure~\ref{Xifig}, Figure~\ref{Pkfig}
shows the $\mu^2$ angular component of the observable power spectra,
given by \beq P_{\mu^2}(k) = 2 P_{\delta}(k)\, \left[\beta +
\frac{x_{\alpha}} {\tilde{x}_ {\rm tot}} \tilde{W}(k) \right]\ , 
\label{eq:Pmu2} \eeq and $P_{\rm un-\delta}(k)$ which equals the 
Poisson power spectrum except for a factor of $(x_{\alpha}/
\tilde{x}_{\rm tot})^2$. In addition to these quantities, $P_{\mu^0}$
yields the baryonic density power spectrum at each redshift, and its
normalization (measured in mK) thus yields the mean brightness
temperature $T_b$. Note that the actual plotted quantity is $|T_b|
[k^3 P(k)/(2 \pi^2)]^{1/2}$, i.e., the square root of the contribution
per $\log k$ of power on the scale $k$ to the variance of the
brightness temperature. The figure shows that $\beta$ can be
determined from $P_{\mu^2}/P_{\mu^4}$ on small scales ($k \ga 1$
Mpc$^{-1}$), making the two cases (adiabatic: $\beta \la 1/3$ or
pre-heated: $\beta \approx 1$) easily distinguishable. As a result,
the mean brightness temperature determines $x_{\rm tot}$ in these two
limits [see eq.~(\ref{eq:combo})]. Measurements at $k \ga 100$
Mpc$^{-1}$ can also independently probe $T_k$ because of the smoothing
effects of the gas pressure and the thermal width of the 21cm
line. 

Note that the value of $x_{\alpha}$ (which dominates $x_{\rm tot}$ in
the cases shown in Figure~\ref{Pkfig}) is interesting in itself, since
it fixes the product of the fraction of gas in galaxies, times the
typical efficiency with which gas is turned to stars and the stars
produce ultra-violet photons inside these galaxies. The Figure also
shows that the power spectra are strongly sensitive to the radial
distribution of flux seen by the gas, which in turns probes the
contribution of photons from the higher-level Lyman series, a
contribution which depends on the stellar IMF and the spectrum of the
sources. In particular, since $\xi_P(l)$ drops to zero at large $l$,
the corresponding Poisson-induced power spectrum approaches a constant
at small $k$, and the plotted quantity is $\propto k^{3/2}$.

X-ray heating of the cosmic gas above the CMB temperature is likely to
occur well before the universe is ionized because it requires an
amount of energy of only $\sim 5\times 10^{-3}[(1+z)/20]$ eV per
baryon, while reionization requires $\sim 13.6 f_{\rm esc}^{-1}$ eV
per hydrogen atom (or even more if recombinations are included). The
X-rays may be produced by stellar winds, supernovae, X-ray binaries,
quasars, or collapsed halos [see \citet{Madau} and \S 2.3.2 in
\citet{WL04a}]. As long as more than $\sim 0.1\%$ of the energy
radiated by galaxies above the Lyman-limit couples to the IGM through
X-ray scattering, $T_k$ exceeds $T_\gamma$ before the completion of
reionization. It is possible, however, that
\Lya coupling between $T_s$ and $T_k$ is achieved before X-rays heat
$T_k$ above $T_\gamma$.

We note that a significant gas fraction may already lie in virialized
minihalos at high redshift. Although this gas may be dispersed due to
internal feedback from star formation due to molecular hydrogen
cooling \citep{ohh, ricotti}, if stars cannot form then the dense gas
minihalos produce 21cm fluctuations \citep{iliev1, iliev2}. Since the
gas content is strongly suppressed in halos below the filtering mass
\citep{gnedin}, we estimate the maximum gas fraction in minihalos as
lying between the filtering mass $M_{\rm F}$ and the minimum mass
needed for atomic cooling. At $z=20$, only around $0.6\%$ of the gas
is in minihalos in the mass range $8 \times 10^5$ -- $3
\times 10^7 M_{\odot}$, compared to $11\%$ in minihalos in the range
$4 \times 10^5$ -- $1 \times 10^8 M_{\odot}$ at $z=8$. Thus, in the
redshift range that we focus on in this paper, only a minor fraction
of the gas is in virialized regions and so we consider the density and
temperature distributions only in uncollapsed regions.

\section{Studying the Earliest Galaxies}
\label{sec:galaxies}

Observing the enhanced 21cm fluctuations would constitute a direct
detection of the collective emission of the earliest generation of
galaxies, long before reionization. Such a detection is potentially
feasible within a few years (http://space.mit.edu/eor-workshop/),
while the number of stars in such an early galaxy may be smaller than
in a present-day globular cluster, making these galaxies at $z\sim
20$--30 undetectable individually even with the proposed James Webb
Space Telescope (http://www.jwst.nasa.gov/) planned for launch in the
next decade.

\begin{figure}
\plotone{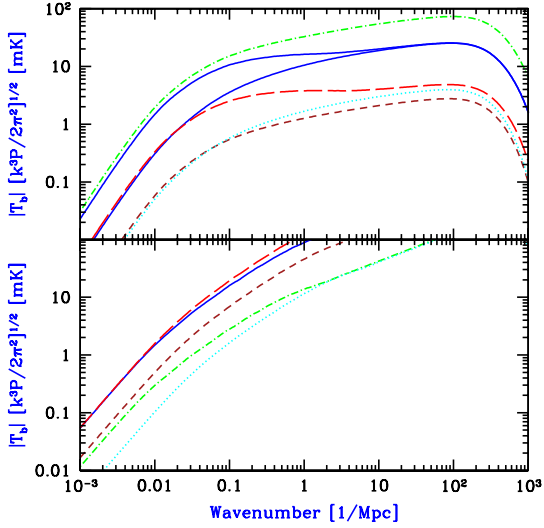}
\caption{Power spectra of 21cm brightness fluctuations versus comoving 
wavenumber. Shown are $P_{\mu^2}$ (top panel), which contains the
contribution of the density-induced fluctuations in flux, and $P_{\rm
un-\delta}$ (bottom panel), which is solely due to the Poisson-induced
fluctuations. We assume atomic cooling with an efficiency that varies
according to local observations (see text), normalized so that $x_{\rm
tot}=1$ at $z=20$. Redshifts shown are $z=35$ (dotted), 30
(short-dashed), 25 (long-dashed), 20 (upper solid), and 15
(dot-dashed). We also show $2
\beta P_{\delta}$ at $z=20$ (top panel, lower solid curve). We 
assume full Lyman-band emission (see \S~\ref{sec:basic}) and Pop III
stars.} \label{zfig1} \end{figure}

In order to illustrate how the properties of the first galaxies may be
probed with 21cm fluctuations, we predict the observed power spectra
at a sequence of redshifts, for various choices of the parameters of
galactic halos and of IGM heating. Figure~\ref{zfig1} shows a
reasonable lower limit on the number of halos in the universe. Here we
maximize the dominance of massive halos by assuming that: {\it (i)}
atomic cooling is required for efficient star formation, and {\it
(ii)} the star formation efficiency declines with circular velocity
according to local observations \citep{Dekel,Kauffman}, emphasizing
the role of high-mass halos. The star formation efficiency is set to
$12\%$ above $V_c = 180~{\rm km~s^{-1}}$ and $\propto V_c^2$ below
this value. In this case, reionization completes at $z=11.7$. In
Figure~\ref{zfig1}, the threshold parameter $x_{\rm tot}$ (going from
high redshift to low) is 0.070, 0.037, 0.12, 1.0, and 6.2. The
corresponding value of $\beta$ is 2.1, 2.0, 0.36, 0.22, and 0.25,
respectively. We find at $z=20$ a filtering mass $M_{\rm F} = 8.0
\times 10^5 M_{\odot}$, $R_{\rm F} = 2.7$ kpc, and $R_{\rm T} = 1.4$
kpc. Collisions dominate the \Lya coupling at $z
\ga 30$ (contributing $x_c=0.070$, 0.029, and 0.008 at the three 
highest redshifts shown), producing a high $\beta$ and suppressing the
effect of the fluctuations in \Lya flux which are largest at the
highest redshifts. The absolute signal in mK tends to rise with time
due to the growth of density fluctuations (which affects $P_{\mu^2}$)
and the declining gas temperature which increases the overall 21cm
absorption in the adiabatic case. However, the increasing number
density of sources with time eventually suppresses the Poisson
fluctuations. Also, the saturation of the \Lya coupling (i.e., $x_{\rm
tot} \gg 1$) reduces the sensitivity of the 21cm fluctuations to
fluctuations in the \Lya flux; in the figure, by $z=15$ the power
spectrum $P_{\mu^2}(k)$ has essentially reverted back to the shape of
$P_{\delta}(k)$. The figure does not include the signature of ionized
bubbles, which becomes significant only at later times when the
ionized fraction approaches unity.

\begin{figure}
\plotone{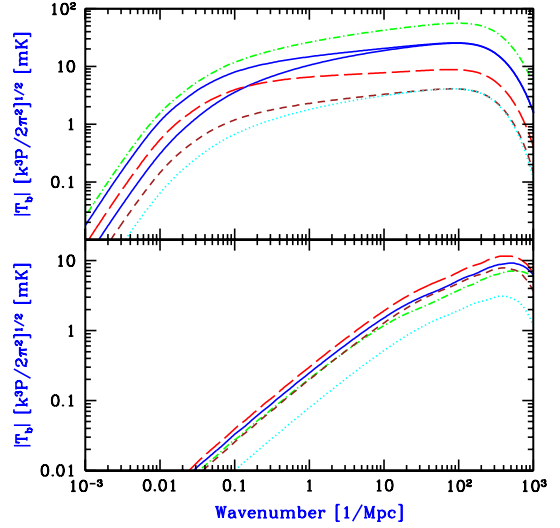}
\caption{Power spectra of 21cm brightness fluctuations versus comoving 
wavenumber. Shown are $P_{\mu^2}$ (top panel), which contains the
contribution of the density-induced fluctuations in flux, and $P_{\rm
un-\delta}$ (bottom panel), which is solely due to the Poisson-induced
fluctuations. We assume H$_2$ cooling with a fixed efficiency set so
that $x_{\rm tot}=1$ at $z=20$. Redshifts shown are $z=35$ (dotted),
30 (short-dashed), 25 (long-dashed), 20 (upper solid), and 15
(dot-dashed).  Note the different axes ranges compared to
Figure~\ref{zfig1}. We also show $2 \beta P_{\delta}$ at $z=20$ (top
panel, lower solid curve). We assume full Lyman-band emission (see
\S~\ref{sec:basic}) and Pop III stars.}
\label{zfig2} \end{figure}

Figure~\ref{zfig2} shows a reasonable upper limit on the number of
halos in the universe. In this case we maximize the contribution of
numerous low-mass halos by assuming that only molecular cooling is
required for efficient star formation, with a fixed star formation
efficiency ($f_*=0.03\%$). With these parameters, the universe would
not complete reionization by redshift 6 without a significant rise in
$f_*$ at $z\ga 6$. Figure~\ref{zfig2} shows some of the same trends
that are apparent in the previous figure, although both sources of
fluctuations in flux are significantly smaller, since there are more
halos contributing and they are more weakly biased. The threshold
parameter $x_{\rm tot}$ (going from high redshift to low) is .076,
.082, 0.30, 1.0, and 1.9. As before, collisions dominate at high
redshift, contributing $x_c=0.070$, 0.029, and 0.008 at the three
highest redshifts shown. The values of $\beta$ are 1.9, 0.95, 0.24,
0.22, and 0.25, respectively.

\begin{figure}
\plotone{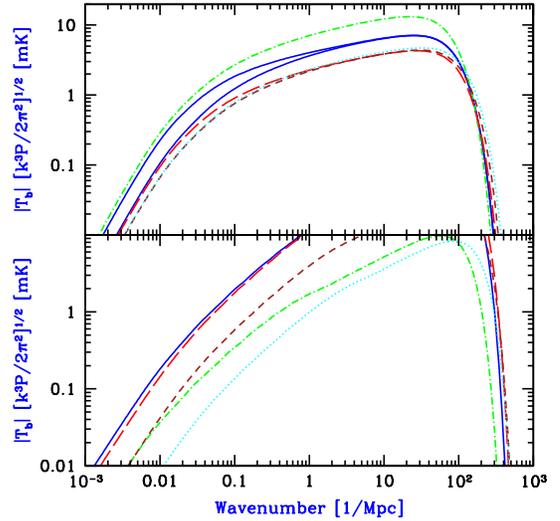}
\caption{Power spectra of 21cm brightness fluctuations versus comoving 
wavenumber. Same as Figure~\ref{zfig1}, except that we assume
pre-heating to 500 K by X-ray sources. Note the different axes ranges
compared to Figure~\ref{zfig1}.} \label{zfig3}
\end{figure}

In Figure~\ref{zfig3} we consider the same halo population as in
Figure~\ref{zfig1}, but with the 21cm fluctuations modified due to an
assumed pre-heating of the IGM. We set the star formation efficiency
to $6.9\%$ above $V_c = 180~ {\rm km~s^{-1}}$ and $\propto V_c^2$
below this threshold, so that $x_{\rm tot}=1$ at $z=20$. The values of
$x_{\alpha}$ at the various redshifts are 0.0003, 0.006, 0.08, 0.79,
and 5.3, while $x_c=0.61$, 0.46, 0.32, 0.21, and 0.12,
respectively. In this case, collisions are still significant at the
lower redshifts because of the increased collision rate at high
temperature. The value of $\beta$ is 1.6, 1.7, 1.6, 1.1, and 1.0,
respectively, always $\ga 1$ since the gas is uniformly heated and no
longer changes its temperature adiabatically. In this case $M_{\rm F}
= 2.4 \times 10^6 M_{\odot}$, $R_{\rm F} = 3.9$ kpc, and $R_{\rm T} =
10.8$ kpc at $z=20$. Comparing Figures~\ref{zfig1} and
\ref{zfig3}, we see that the strong differences between these two
cases in both the mean $T_b$ and in the value of $\beta$ (which
affects $P_{\mu^2}$) allow for a consistency check on the thermal
state of the IGM. Note that in the cases shown in Figures~\ref{zfig1}
and \ref{zfig3}, the Poisson fluctuations diverge on small scales, in
which case our linear analysis based on equation~(\ref{powTb})
represents an estimate that is subject to strong non-linear
corrections. Also, on scales $\la 0.1$ comoving Mpc, which corresponds
to the initial region that collapses to become a typical halo at $z
\sim 20$, our calculation is subject to corrections due to non-linear
gravitational collapse.

From the ratio $P_{\mu^2}/P_{\mu^4}$ at large scales, it is possible
to measure $\tilde{W}(0)$ [see eq.~({\ref{eq:Pmu2})], which yields
directly the average bias factor of the emitting sources, given by
equation~(\ref{eq:Wk}) with the curly brackets becoming simply $\{1 +
b(z')\}$ in this limit. The relative bias of galaxies is on large
scales predicted to depend simply on the rarity (\# of sigma) of the
underlying Gaussian density fluctuations that led to their
formation. Thus, it is sensitive to the typical mass of the dark
matter halos that host galaxies at these redshifts.  This large-scale
enhancement factor can be easily seen by comparing the upper solid
line to the dashed line in Figure~\ref{Pkfig}, and the pair of $z=20$
lines in each of Figures~\ref{zfig1}, \ref{zfig2}, and \ref{zfig3}. 
The enhancement factor shown in the figure (which equals the square
root of the ratio between $P_{\mu^2}$ and $2 \beta P_{\delta}$) is in
each case 4.8, 5.1, 4.0, and 2.2, respectively. The corresponding
value of $\tilde{W}(0)$ is 9.8, 11.2, 6.5, and 11.2, respectively.

\section{Summary}

We have shown that the 21cm flux fluctuations from intergalactic
hydrogen are strongly enhanced due to the radiation emitted by the
first galaxies at redshifts $\sim 20$--$30$.  The earliest galaxies
emitted photons with energies between the Ly$\alpha$ and Lyman-limit
transitions of hydrogen, to which the neutral universe was transparent
except at the Lyman series resonances. As these photons redshifted
into a Lyman-series resonance and then degraded into a Ly$\alpha$
photon, they coupled the spin temperature $T_s$ of the 21cm transition
of hydrogen to the gas kinetic temperature $T_k$, allowing it to
deviate from the microwave background temperature, $T_\gamma$. Figures
4--7 show that the fluctuations in the radiation emitted by the first
galaxies enhanced the overall power-spectrum of 21cm flux fluctuations
by at least a factor of a few during the period when \Lya coupling was
established. The enhancement is sourced by inhomogeneities in the
biased distribution of galaxies (as those represent rare peaks in the
density field) as well as by Poisson noise in the number of galaxies.
The effects of galaxies are more apparent when the power spectrum is
split into angular components relative to the line of sight (with the
anisotropy sourced by peculiar velocities); in particular, the
component $P_{\rm un-\delta}$ shown in the lower panels of Figures
4--7 (eq.~\ref{eq:un-d}) would vanish in the absence of Poisson
fluctuations in the distribution of galaxies. For the density-induced
fluctuations (upper panels), the enhancement factor is roughly
constant on scales $\ga 10$ comoving Mpc. Because of the rapid
evolution in the abundance of galaxies at early cosmic times, and the
atomic physics of resonant scattering by hydrogen, the fluctuations in
the \Lya intensity are dominated by the contribution from sources
within distances $\sim 1$--100 comoving Mpc, well within the source
horizon out to which a \Lyb photon could redshift into a \Lya photon
(Fig.~\ref{radialfig}).

Several experiments are currently being constructed or designed to
detect the 21cm fluctuations prior to reionization. For an overview,
see http://space.mit.edu/eor-workshop/; for individual experiments,
see LOFAR (http://www.lofar.org/), MWA
(http://web.haystack.mit.edu/arrays/MWA/index.html), PAST
(http://xxx.lanl.gov/abs/astro-ph/0404083), and the future SKA
(http://www.skatelescope.org/). Although low-frequency foregrounds are
much brighter than the 21cm signal, they are not expected to include
sharp spectral features, unlike the 21cm maps that should decorrelate
over small shifts in frequency (corresponding to slicing the universe
at different redshifts). Consequently, the prospects for extraction of
the 21cm signal are promising \citep{Shaver, Zalda04, Miguel1,
Miguel2, Santos}. The signal-to-noise ratio in these experiments is
controlled by the sky brightness and grows as the square-root of the
integration time.  Hence, the increase in the amplitude of 21cm
fluctuations that we predict has important practical implications for
observing the 21cm signal from the corresponding redshifts, since the
required integration time is reduced by the square of the enhancement
factor.

The enhanced level of fluctuations in the 21cm flux lasted only until
the \Lya coupling of $T_s$ and $T_k$ was saturated. Both before and
after the \Lya coupling epoch, the enhancement (compared to the case
of a uniform \Lya flux) had a lower amplitude (see, e.g.,
Figure~\ref{zfig1} where \Lya coupling was achieved at $z\sim
20$). Detection of the enhancement factor over the available redshift
interval (i.e., the corresponding observed frequency band) can be used
to infer the abundance and characteristic mass of the earliest
galaxies (\S 7) without any such galaxy being observed
individually. The existence of
\Lya coupling can also be used to determine whether $T_k$ is larger
than $T_\gamma$ in this redshift interval. If the 21cm signal appears
in emission rather than in absorption [where absorption is present in
the interval $30\la z\la 200$ \citep{Loeb04}] then $T_k>T_\gamma$,
most likely due to heating from X-rays emitted by the same galaxies.
Altogether, the 21cm fluctuations can be used to probe the level of UV
and X-ray radiation emitted by the earliest galaxies, long before
reionization was complete. The milestones of \Lya coupling between
$T_s$ and $T_k$ and X-ray heating of $T_k$ above $T_\gamma$ provide
important precursors for the radiative feedback of galaxies on the IGM
that culminated with the completion of reionization.

\acknowledgments

We acknowledge support by NSF grant AST-0204514. R.B. is grateful for
the kind hospitality of the Harvard-Smithsonian CfA, and acknowledges
the support of NSF grant PHY99-07949 at KITP, Israel Science
Foundation grant 28/02/01, and an Alon Fellowship at Tel Aviv
University. This work was also supported in part by NASA grant NAG
5-13292 and NSF grants AST-0071019 (for A.L.).

\end{document}